\documentclass[superscriptaddress,pra,aps,10pt,longbibliography,notitlepage]{revtex4-1}

\usepackage{amsfonts}
\usepackage{amssymb}
\usepackage{amsmath}
\usepackage{amsthm}
\usepackage{newfloat,algcompatible}
\usepackage[size=small]{caption}
\usepackage{etoolbox}
\AtBeginEnvironment{algorithm}{\noindent\par\nobreak\vskip-5pt}
\usepackage{newfloat}
\selectfont{}

\DeclareFloatingEnvironment[
    fileext=loa,
    listname=List of Algorithms,
    name=ALGORITHM,
    placement=tbhp,
]{algorithm}
\DeclareCaptionFormat{algorithms}{\vskip-15pt\hrulefill\par#1#2#3\vskip-6pt\hrulefill}
\usepackage{graphicx}
\usepackage{verbatim}
\usepackage{hyperref}
\usepackage{bbm}
\usepackage[english]{babel}
\usepackage[normalem]{ulem}
\usepackage{natbib}
\usepackage[pdftex]{color}
\usepackage[dvipsnames]{xcolor}
\usepackage{braket}
\usepackage[utf8]{inputenc}
\usepackage{enumitem}
\setlist[description]{leftmargin=\parindent,labelindent=\parindent}

\newcommand{\Prob}[1]{\ensuremath{P_{\textnormal{#1}}}}
\newcommand{\Psame}{\Prob{same}}
\newcommand{\Pboth}{\Prob{both}}
\newcommand{\Pone}{\Prob{one}}
\newcommand{\Pneither}{\Prob{neither}}
\newcommand{\Pdif}{\Prob{dif}}
\newcommand*\cp{c^\prime}

\newtheorem{definition}{Definition}

\graphicspath{{figs/}}
\hypersetup{
    colorlinks=true,
    linkcolor=cyan,
    citecolor=magenta,
    filecolor=magenta,
    urlcolor=cyan,
    runcolor=cyan
}
\begin{document}

\title{A Parameter Setting Heuristic for the Quantum Alternating Operator Ansatz}

 \author{James Sud}
 \email{jsud@uchicago.edu}
 \thanks{Majority completed at USRA}
 \affiliation{Quantum Artificial Intelligence Laboratory (QuAIL), NASA Ames Research Center, Moffett Field, CA, 94035, USA}
 \affiliation{USRA Research Institute for Advanced Computer Science (RIACS), Mountain View, CA, 94043, USA}
 \affiliation{Department of Computer Science, University of Chicago, 5730 S Ellis Ave, Chicago, IL 60637, USA}

 \author{Stuart Hadfield}
 \affiliation{Quantum Artificial Intelligence Laboratory (QuAIL), NASA Ames Research Center, Moffett Field, CA, 94035, USA}
 \affiliation{USRA Research Institute for Advanced Computer Science (RIACS), Mountain View, CA, 94043, USA}
 
 \author{Eleanor Rieffel}
 \affiliation{Quantum Artificial Intelligence Laboratory (QuAIL), NASA Ames Research Center, Moffett Field, CA, 94035, USA}
 
 \author{Norm Tubman}
 \affiliation{Quantum Artificial Intelligence Laboratory (QuAIL), NASA Ames Research Center, Moffett Field, CA, 94035, USA}
 
 \author{Tad Hogg}
 \affiliation{Quantum Artificial Intelligence Laboratory (QuAIL), NASA Ames Research Center, Moffett Field, CA, 94035, USA}

\date{\today}

\begin{abstract}

Parameterized quantum circuits are widely studied approaches to tackling challenging optimization problems. A prominent example is the Quantum Alternating Operator Ansatz (QAOA), a generalized approach that builds on the alternating structure of the Quantum Approximate Optimization Algorithm. Finding high-quality parameters efficiently for QAOA remains a major challenge in practice. In this work, we introduce a classical strategy for parameter setting, suitable for cases in which the number of distinct cost values grows only polynomially with the problem size, such as is common for constraint satisfaction problems. 
The crux of our strategy is that we replace the cost function expectation value step of QAOA with a classical model that also has parameters which play an analogous role to the QAOA parameters, but can be efficiently evaluated classically. This model is based on empirical observations of QAOA, where variable configurations with the same cost have the same amplitudes from step to step, which we define as Perfect Homogeneity. Perfect Homogeneity is known to hold exactly for problems with particular symmetries. More generally, high overlaps between QAOA states and states with Perfect Homogeneity have been empirically observed in a number of settings. Building on this idea, we define a Classical Homogeneous Proxy for QAOA in which this property holds exactly, and which yields information describing both states and expectation values. We classically determine high-quality parameters for this proxy, and then use these parameters in QAOA, an approach we label the Homogeneous Heuristic for Parameter Setting. We numerically examine this heuristic for MaxCut on unweighted Erd\H{o}s-R\'{e}nyi random graphs. For up to $3$ QAOA levels we demonstrate that the heuristic is easily able to find parameters that match approximation ratios corresponding to previously-found globally optimized approaches. For levels up to $20$ we obtain parameters using our heuristic with approximation ratios monotonically increasing with depth, while a strategy that uses parameter transfer instead fails to converge with comparable classical resources. These results suggest that our heuristic may find good parameters in regimes that are intractable with noisy intermediate-scale quantum devices. Finally, we outline how our heuristic may be applied to wider classes of problems.

\end{abstract}

\maketitle

\section{Introduction}\label{introduction}

The Quantum Alternating Operator Ansatz (QAOA)~\cite{hadfield19} is a widely-studied parameterized quantum algorithm for tackling combinatorial optimization problems. It uses the alternating structure of Farhi et al.’s Quantum Approximate Optimization Algorithm~\cite{farhi14}, a structure that was also studied in special cases in much earlier work \cite{hogg00, hogg00-1}. Recent work has explored regimes for which suitable parameters for QAOA are difficult to pre-compute. This includes extensive research viewing QAOA as a variational quantum-classical algorithm (VQA), in which results from runs on quantum hardware are fed into a classical outer loop algorithm for updating the parameters. Optimizing parameters for VQAs can be quite challenging, as one typically needs to search over a large parameter space with a complex cost landscape. While for a successful algorithm, one does not necessarily need to find optimal parameters, but rather good enough  parameters for a given target outcome, finding good parameters can be difficult due to the number of local optima \cite{cerezo21, wierichs20}, and in some cases barren plateaus \cite{mcclean18, larocca22}. Moreover, the high levels of noise present on current quantum devices can exacerbate these challenges~\cite{wang21}. 

\begin{figure}[!htb]
    \centering
    \includegraphics[width=0.65\textwidth]{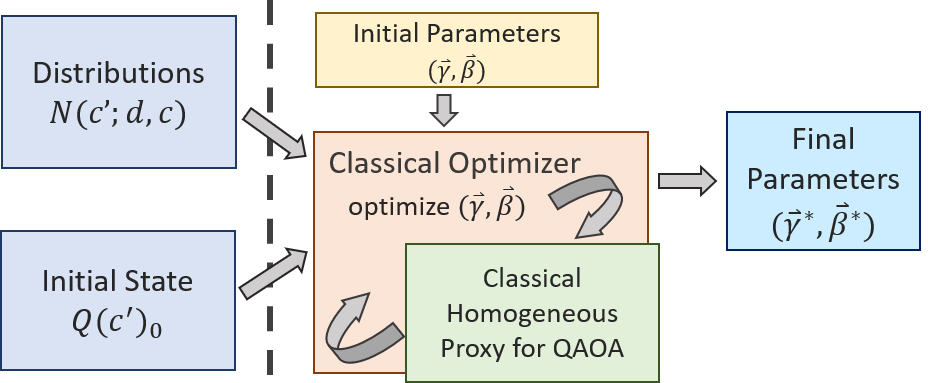}
    \caption{Flowchart of parameter setting procedure using the Classical Homogeneous Proxy for QAOA. The boxes to the left of the dashed line denote inputs that can be viewed as classical pre-computations, as described in Sec.~\ref{homogeneous_approximation}. Given these inputs, along with an initial guess of parameters, one can perform parameter setting, in which a classical optimizer and the proxy are used in a loop to search for parameters for the algorithm. Here, the proxy is used to estimate values of the cost function in order to reduce the time and space complexity for cost function evaluation, while the choice of classical optimizer and initial parameters is left open. In the end, the procedure outputs good parameters for the proxy that are then used as parameters for QAOA.}
    \label{fig:flowchart}
\end{figure}

We propose a novel approach to QAOA parameter optimization that maps the QAOA circuit to a simpler classical model. The crux of our approach is that in the parameter objective function (as introduced below), we replace the cost function expectation value, which is typically evaluated either on quantum hardware or using expensive classical evaluation, with a simpler model that has parameters that play an analogous role to the QAOA parameters, but can be efficiently evaluated classically. This approach is based on the observation, originally due to Hogg \cite{hogg00, hogg00-1}, that one may leverage a classical \lq\lq mean-field\rq\rq\ or \lq\lq homogeneous\rq\rq\ model where variable configurations with the same cost have the same amplitudes from step to step, which we define as Perfect Homogeneity. Extending this idea, we define a Classical Homogeneous Proxy for QAOA (``proxy'' for short) in which this property holds exactly, and which yields both information describing states and expectation values. We then classically determine high-quality parameters for this proxy, and then use these parameters in QAOA, an approach we label the Homogeneous Heuristic for Parameter Setting (``heuristic'' for short). This heuristic is visualized in Fig.~\ref{fig:flowchart}. 

The heuristic is appropriate for classes of constraint satisfaction problems (CSPs) in which instances consists of a number of clauses polynomial in the number of variables, with each clause acting on a subset of the variables. For such problems, the number of distinct cost function values, and thus the computation time of the proxy, is polynomially bounded in the number of variable as desired. We describe the proxy for four such classes of CSPs: random kSAT, MaxCut on unweighted Erd\H{o}s-R\'{e}nyi model graphs, random MaxEkLin2, and random Max-k-XOR. For these examples, the proxy leverages information only about the class (as well as the number of variables and clauses), rather than problem instance, so that the proxy describes states and expectation values for the entire class. We then explore the heuristic for parameter setting on classes of MaxCut problems. This heuristic returns a set of parameters for the entire \textit{class}, which can then be tested on instances from that class. Our results indicate that for MaxCut, the heuristic on $20$-node graphs is able to -- for $p=1$ ,$2$, and $3$ -- identify parameters yielding approximation ratios comparable to those returned by the commonly-used strategy of transferring globally-optimized parameters~\cite{lotshaw21,shaydulin21_qaoakit}. We then demonstrate out to depth $p=20$ that the heuristic identifies parameters yielding monotonically increasing approximation ratios as the depth of the algorithm increases. A parameter setting strategy that uses an identical outer loop parameter update scheme but does not leverage the proxy fails to converge given identical classical resources. 

The practical implications of this work is that for classes of optimization problems such as CSPs with a fixed number of randomly selected clauses on a fixed number of variables, the Classical Homogeneous Proxy for QAOA can be computed with solely classical resources in time polynomially scaling with respect to $p$ as well as $n$. Thus, the proxy avoids the issue of exponentially growing resources (with respect to $p$ or $n$) of using exact classical evaluation of expectation values, and avoids the noise present on NISQ devices. The Homogeneous Heuristic for Parameter Setting leverages this proxy, so the subroutine evaluating cost expectation values may be much quicker. However, the parameter landscape may still be challenging to optimize over, especially in cases in which the number of free parameters grows with the problem size. Nevertheless, we demonstrate for MaxCut that our heuristic is able to outperform previous results of parameter transfer \cite{lotshaw21}, indicating the potential of our heuristic approach (which may also be combined with other advancements in parameter setting, as discussed briefly later in the paper) to achieve further improvements in practice.

\emph{Quantum Alternating Operator Ansatz}: We here briefly describe the structure of QAOA \cite{farhi14,hogg00, hogg00-1,hadfield19}. Given a cost function $c(x)$ to optimize, a QAOA circuit consists of $p$ repeated applications of a mixing and cost Hamiltonian $B$ and $C$ in alternation, parameterized by $2p$ parameters $(\vec{\gamma}, \vec{\beta})$:
\begin{equation}\label{eq:qaoa_ansatz}
    \ket{\Psi(\vec{\gamma}, \vec{\beta})} = e^{-i\beta_pB}e^{-i\gamma_pC}\cdots e^{-i\beta_1B}e^{-i\gamma_1C} \ket{\psi_0},
\end{equation}
where $\ket{\psi_0}$ is an easily-prepared initial state. There is freedom in choosing the Hamiltonians $B$, and $C$, although typically (as is followed in this work), $C$ is chosen such that $C\ket{x}=c(x)\ket{x}$ for all $x$, and $B$ is a mixer that is simple to implement on hardware. More general operators and initial states for QAOA are proposed in \cite{hadfield19}. In this work we choose the X mixer $B = \sum_{i=1}^n X_i$, where $n$ represents the total number of qubits in the circuit, and the starting state $\ket{\psi_0}$ is chosen to be the uniform superposition over all $2^n$ bitstrings, as originally considered in \cite{farhi14}. The goal of QAOA is then to produce a state $\Psi(\vec{\gamma}, \vec{\beta})$ such that repeated sampling in the computational basis yields either an optimal or high-quality approximate solution to the problem. Finding such good parameters is the parameter setting problem and may be approached in a number of ways with different tradeoffs, ranging from black-box optimization techniques to application specific approaches. We refer to fixed QAOA parameters $p,\vec{\gamma},\vec{\beta}$ as a \textit{parameter schedule}.

We now describe QAOA as a VQA: given some classical cost function $c(x)$ on $n$ variable bitstrings $\{0,1\}^n$, and a quantum circuit ansatz\"e with parameters $\vec{\Theta}$, one defines a parameter objective function $f(\vec{\Theta})$ and optimizes over $\vec{\Theta}$ with respect to $f$. In order to optimize over $\vec{\Theta}$, a two-part cycle is typically employed. First, a classical outer loop chooses one or more parameters $\vec{\Theta}$ for which evaluations of $f(\vec{\Theta})$ or its partial derivatives are requested, based on initial or prior cycle information. Then, a subroutine evaluates $f$ and its derivatives for all $\vec{\Theta}$ requested by the outer loop. Given these evaluations, the parameter update scheme can then update the current best $\vec{\Theta}$ and choose a new set of parameters to test. Throughout this work we refer to the outer loop as the parameter update scheme and the inner subroutine as parameter objective function evaluation. Typically in QAOA, $f(\vec{\Theta})= f(\vec{\gamma},\vec{\beta})$ is taken to be $\braket{\vec{\gamma},\vec{\beta}|C|\vec{\gamma},\vec{\beta}}$. This choice of $f$ measures the expected cost function value obtained from sampling the QAOA state, which we refer to as the typical parameter objective function. In this work, we will replace it with what we define as the homogeneous parameter objective function, which utilizes the Classical Homogeneous Proxy for QAOA.

\emph{Related Work}: There have been numerous parameter setting strategies proposed for QAOA. Most of these strategies focus on improving the parameter update scheme, while keeping the typical parameter objective function. These strategies range from the simplest approach of directly applying classical optimizers to approaches incorporating more sophisticated machine learning techniques \cite{khairy20, zhou20, crooks18, shaydulin19, alam2020accelerating, verdon19, wilson21,  skolik2021layerwise, rivera2021avoiding}. In practice, however, efficiently finding high-quality parameters remains a challenging task. Global optimization strategies often get stuck in navigating the parameter optimization landscape due to barren plateaus \cite{mcclean18, wang21, larocca22} or multitudes of local optima \cite{cerezo21, wierichs20}, and the problem becomes especially challenging as the number of parameters grows due to the curse of dimensionality. In some circumstances parameter optimization can even become NP-hard~\cite{bittel2021training}. There have been multiple methods proposed that attempt to alleviate these issues. The first of these include initializing parameters to be close to parameters informed by previous information or intuition \cite{streif20, shaydulin21_symmetries, brady21}. One such example is to use parameters that represent a discretized linear quantum annealing schedule \cite{sack21, wurtz21_counteradiabaticity}. Another example is based on the principle that globally optimal parameters for random QAOA instances drawn from the same distribution tend to concentrate \cite{brandao18, farhi14, zhou20, lee21}. Thus, if globally optimal parameters are known for one, or many instances of a specific class of problems, these exact parameters (or median parameters for more than one instance) empirically perform well on a new, unseen instance from the same class \cite{galda21, shaydulin21_qaoakit}. Another approach for improving parameter setting is re-parameterizing the circuit, which places constraints on the allowed parameter schedules, thereby reducing the number of parameters so that they are optimized more easily. In some cases, this re-parameterization can be informed by optimized parameters for typical instances, or by connections to quantum annealing \cite{zhou20, brady2021behavior, brady21, shaydulin21_symmetries, wurtz21_counteradiabaticity}. 

An alternative paradigm for parameter setting are methods that modify the parameter objective function itself. Indeed, one can find closed form expressions for expected cost as a function of parameters in some cases, for example MaxCut at $p=1$ \cite{wang18}, or $p=2$ for high-girth regular graphs \cite{marwaha21}. Moreover, when applicable one can take advantage of problem locality considerations (such as graphs of bounded vertex degree for MaxCut) to compute the necessary expectation values without requiring the full quantum state vector~\cite{farhi14,wang18}. In these cases, then, one could optimize parameters with respect to these expressions rather than by evaluating the entire dynamics. Other examples of parameter objective function modification include using conditional value at risk \cite{barkoutsos20} and Gibbs-like functions \cite{li20}, which are closely related to the usual parameter objective function. Similar in spirit to our approach, recent work~\cite{sung2020using,shaffer22,mueller2022accelerating} has proposed the use of surrogate models, which use quantum circuit measurement outcomes to construct an approximate parameter objective function. In contrast, our approach is a entirely classical, and the parameters it outputs may be used directly, or possibly improved further, given access to a quantum device. Additionally, a related perspective was recently proposed in \cite{DiezValle22}, wherein the authors study the connection between single-layer QAOA states and pseudo-Boltzmann states where computational basis amplitudes are also expressed as functions of the corresponding costs, i.e., perfectly homogeneous in our terminology. While~\cite{DiezValle22} provides additional motivation for our approach, the authors there do not consider cases beyond $p=1$ and so their results do not immediately apply to parameter setting in the same way. 

\emph{Outline of paper}: In Sec.~\ref{homogeneous_approximation} we define the Classical Homogeneous Proxy for QAOA. In Sec.~\ref{n_dists_random}, we demonstrate how to implement the proxy for classes of CSPs with a fixed, polynomial number of randomly drawn clauses. In Sec.~\ref{homogeneous_parameter_setting} we introduce the Homogeneous Heuristic for Parameter Setting. In Sec.~\ref{validity_of_approximation} we provide numerical evidence for the efficacy of the proxy and the heuristic applied to MaxCut on unweighted Erd\H{o}s-R\'{e}nyi model graphs. Finally in Sec.~\ref{results} we present the results of the heuristic for the MaxCut on the same class of graphs. We conclude in Sec.~\ref{discussion} with a discussion of our results and several future research directions.

\section{Classical Homogeneous Proxy for QAOA}\label{homogeneous_approximation}

This section summarizes and generalizes the approach of \cite{hogg00}, updating the notation to match current notation for QAOA and allowing for easier extension to other CSPs. In this section, we describe  the Classical Homogeneous Proxy for QAOA from a sum-of-paths perspective~\cite{hogg00,hadfield2021analytical} using a set of classical equations that ensure what we presently define as Perfect Homogeneity.

\begin{definition}\label{def:perfect_homogeneity}
\textbf{Perfect Homogeneity}: For a given state $\ket{\Psi}=\sum_{x=\{0,1\}^n} q(x) \ket{x}$, where $q(x)$ is the amplitude of bitstring $x$ when written in the computational basis, $\ket{\Psi}$ has Perfect Homogeneity if and only if the the amplitude $q(x)$ of all $x$ can be written solely as a function of $c(x)$, i.e. $Q(c(x))$. Then $\ket{\Psi}=\sum_{x={0,1}^n} Q(c(x)) \ket{x}$.
\end{definition}

Perfect Homogeneity trivially holds for QAOA states with non-degenerate cost functions where each bitstring $x$ has a unique cost $c(x)$, as well as the maximally degenerate case where the cost function is constant. A less trivial example is the class of cost functions that depend only on the Hamming weight of each bitstring, $c(x)=c(|x|)$, such as the Hamming ramp $c(x) \sim |x|$. For this class, symmetry arguments~\cite{shaydulin21_symmetries} imply that the QAOA state retains Perfectly Homogeneity for any choice of QAOA circuit depth and parameters. Indeed, useful intuition for homogeneity can be gained from the symmetry perspective; in \cite[Lem. 1]{shaydulin21_symmetries} it is shown that for a classical cost function with symmetries compatible with the QAOA mixer and initial state, bitstrings that are connected by these symmetries will have identical QAOA amplitudes. In this case the corresponding QAOA states belong to a Hilbert space of reduced dimension. Perfect Homogeneity is an even stronger condition: all bitstrings with the \emph{same cost} have identical amplitudes, not just those connected by mixer-compatible symmetries. Hence, QAOA states can be expected to be Perfectly Homogeneous only in limited cases, though these states may be near to Perfectly Homogeneous states in some settings~\cite{hogg00}. We consider QAOA applied to cost functions with polynomially many distinct cost values, such that states may not have Perfect Homogeneity. Classically, however, we may assert Perfect Homogeneity to construct our proxy. 

Additional intuition comes from the case of strictly $k$-local problem Hamiltonians (such as, for instance, MaxCut), where the QAOA state is easily shown to be Perfectly Homogeneous to leading-order in $|\gamma|$, independent of $\beta$, with amplitude of bitstring $x$ given by $\frac{1}{\sqrt{2^n}} ( 1-i\gamma c(x) e^{i 2k})$ in the $p=1$ case. Similar analysis for the higher $p$ case also yields Perfect Homogeneity to first order in the $\gamma_j$ parameters. This connection is considered further in Sec.~\ref{numerical_overlaps}.

Given this intuition, we begin from the sum-of-paths perspective~\cite{hadfield2021analytical} for QAOA to provide preliminary analysis for our approach.

\subsection{Classical Homogeneous Proxy for QAOA 
from the Sum-of-Paths Perspective for QAOA}\label{sum_of_paths_perspective}

The amplitude $q_{\ell}(x)$ of a bitstring $x$ at step $\ell$ induced by applying a layer of QAOA with parameters $(\gamma, \beta)$ to a QAOA state with $\ell-1$ layers can be expressed succinctly as~\cite{hadfield2021analytical}
\begin{equation}\label{eq:general_amp_sop}
    q_{\ell}(x) = \braket{x|\vec{\gamma},\vec{\beta}}=\sum_y q_{\ell-1}(y) \cos^{n-d_{xy}}\beta(-i\sin\beta)^{d_{xy}} e^{-i\gamma c_y}, 
\end{equation}
where $d_{xy}$ is the Hamming distance between bitstrings $x$ and $y$, $\cos\beta^{n-d_{xy}}(-i\sin\beta)^{d_{xy}}=\bra{x}e^{-i\beta B} \ket{y}$ are the mixing operator matrix elements, $c_y$ is the cost of bitstring $y$, and the sum is over all $2^n$ bitstrings $y$ in the computational basis.

Grouping the terms in Eq.~\eqref{eq:general_amp_sop}, we can rewrite the sum as 
\begin{equation}
    q_{\ell}(x) = \sum_{d,c}  \cos^{n-d}\beta(-i\sin\beta)^d e^{-i\gamma c} \sum_{y|d_{xy}=d, c_y=c} q_{\ell-1}(y)
\end{equation}
where the sum over $d$ runs from $[0,n]$ and the sum over $c$ runs over the set of unique costs allowed by the cost function, which we describe in Sec.~\ref{cost_distributions}. If it is the case that the amplitudes $q_{\ell-1}(x)$ of all bitstrings with cost $\cp$ are identical, then we can substitute $q(x)$ with $Q(c(x))$ for all $x$. This is exactly Perfect Homogeneity as described in Def.~\ref{def:perfect_homogeneity}. For the rest of this text we use this upper case $Q(c(x))$ to distinguish a function of cost $c=c(x)$ rather than the bitstring $x$ itself. This substitution yields
\begin{equation}\label{eq:pre_homog_sop}
    q_{\ell}(x) = \sum_{d,c}  \cos^{n-d}\beta(-i\sin\beta)^d e^{-i\gamma c_y} Q_{\ell-1}(c) n(x;d,c),
\end{equation}
where $n(x;d,c)$ represents the number of bitstrings with cost $c$ that are of Hamming distance $d$ from $x$. We note that for this equation, and all following equations, this evolution is no longer restricted to unitary evolution. As such, the values $q$ and $Q$ no longer represent strictly quantum amplitudes, but rather track an object that is an analogue to the quantum amplitude. We now introduce the major modification to the sum-of-paths equations. For this, we replace each distribution $n(x;d,c)$ with some distribution $N(\cp;d,c)$ where $c_x=\cp$, where we again adopt the upper case $N$ to distinguish a new distribution that solely depends on the cost of the bitstring. This distribution will generally differ from the original, but in this work we exhibit cases where $N(\cp;d,c)$ is efficiently computable and can adequately replaces $n(x;d,c)$ for the purpose of parameter setting. These cases are discussed in Sec.~\ref{n_dists_random} and the effectiveness of the replacement is numerically explored in Sec.~\ref{validity_of_approximation}.

Using $N(\cp;d,c)$ distributions to replace $n(x;d,c)$, then, we can further rewrite the sum as
\begin{equation}\label{eq:homog_sop}
    q_{\ell}(x) = Q_{\ell}(\cp) = \sum_{d,c}  \cos^{n-d}\beta(-i\sin\beta)^d e^{-i\gamma c} Q_{\ell-1}(c) N(\cp;d,c),
\end{equation}
where $Q_{\ell}(\cp)$ is used to make explicit that the substitutions yield an expression for $q_{\ell}(x)$ that depends solely on the cost $\cp$ of bitstring $x$. This leads to a crucial point for our analysis: if we start in a state observing Perfect Homogeneity, and if the distributions $n(x;d,c)$ can be replaced by distributions $N(\cp;d,c)$, which solely depend on the cost $\cp$ of $x$, then subsequent layers retain the Perfect Homogeneity of the state, yielding analogues of amplitudes $Q(\cp)$. Thus, Eq.~\eqref{eq:homog_sop} is a recursive equation yielding Perfectly Homogeneous analogues of quantum states, which we label the Classical Homogeneous Proxy for QAOA. We here note that importantly, the proxy will not return the analogue of amplitude of some bitstring $x$, but rather the analogue of amplitude of bitstrings with cost $\cp$, as implicit knowledge of which bitstrings $x$ have cost $\cp$ would solve the underlying objective function. In Sec.~\ref{validity_of_approximation} we argue that there are cases where these analogue of amplitudes $Q(\cp)$ are close to all $q(x)$ with $c_x=\cp$. 

\subsection{Cost Distributions}\label{cost_distributions}

In order to evaluate Eq.~\eqref{eq:homog_sop}, we note that the set of costs to which indices $c$ and $\cp$ belong must be defined. Ideally, this set represents exactly the unique objective function values allowed by the underlying cost function. For optimization problems, however, this set is unknown, and is precisely what we would like to solve. Instead, we can form a reasonable estimate by determining upper and lower bounds on the cost function value, as well as by excluding energies which can be efficiently determined to be excluded by the cost function. In this work we denote this estimated set of unique costs as $\mathcal{C}$. As an example, for CSPs with binary valued clauses, the cost function counts the number of satisfied clauses, trivially yielding $\mathcal{C}=\{0,1,...,m\}$, where $m$ is the number of clauses. The set $\mathcal{C}$ also allows us to define the initial state $Q_0(\cp)$ for the algorithm. QAOA typically begins with a uniform superposition over all bitstrings $x$, such that $q_0(x) = \frac{1}{\sqrt{2^n}}$ for all $x$. Thus we can set $Q_0(\cp)=\frac{1}{\sqrt{2^n}}$ for all $\cp$ in $\mathcal{C}$.

While Eq.~\eqref{eq:homog_sop} yields analogues of amplitudes $Q_{\ell}(\cp)$, one may also wish to use the Classical Homogeneous Proxy for QAOA to return an estimate of expected value of the cost function. This requires computing a distribution $P(\cp)$ for all $\cp$ in $\mathcal{C}$, representing the probability a randomly chosen bitstring has cost $\cp$, averaged over the entire class. Much like in the case of computing $\mathcal{C}$, this computation is approximate, as a perfect computation of this distribution would solve the underlying objective function. Examples of estimations of $P(\cp)$ are shown in Sec.~\ref{n_dists_random}. In order to compute our estimate of expected value of the cost, then, we simply sum over costs $\cp$, weighting by the expected number of bitstrings with cost $\cp$ ($2^n P(\cp)$) and the squared analogue to the amplitude $|Q_{\ell}(\cp)|^2$,

\begin{equation}\label{eq:homog_sop_cost}
\widetilde{\braket{C}}= \sum_{\cp}  2^n P(\cp)|Q_{\ell}(\cp)|^2 \cp,
\end{equation}
with the tilde indicating the use of the proxy and that this is no longer a strictly normalized quantum expectation value. This is exactly the equation we use for the homogeneous parameter objective function as described in the introduction. Note that $P(c')$ does not give a bona fide probability distribution unless it is re-normalized by dividing by $\sum_{\cp}  2^n P(\cp)|Q_{\ell}(\cp)|^2$ at each step; nevertheless we have found these factors remain close to unity for the parameter setting experiments considered below and so we neglect them going forward, which yields further computational savings. It is straightforward to introduce these factors into Eq.~\ref{eq:homog_sop_cost} if desired in other applications.

The set $\mathcal{C}$ and estimate to cost $\widetilde{\braket{C}}$ are critically determined by the class of problems one is working with. Examples of these values for a sample of classes is given in Sec.~\ref{n_dists_random}.

\subsection{Algorithm for Computing the Classical Homogeneous Proxy for QAOA}

Formalizing the results in this section, we present Algorithm~\ref{alg:homog_apx_evol}, which describes how given QAOA parameters $p,\vec{\gamma},\vec{\beta}$ a set of possible costs $\mathcal{C}$, and assumed cost distribution $N(\cp;d,c)$ we can compute the Classical Homogeneous Proxy for QAOA using Eq.~\eqref{eq:homog_sop} and Eq.~\ref{eq:homog_sop_cost}.
\vspace{2mm}

\begin{algorithm}
\caption{Classical Homogeneous Proxy for QAOA}\label{alg:homog_apx_evol}
\begin{algorithmic}
\STATE \textbf{Input}: $p$, $\vec{\gamma}$, $\vec{\beta}$, $n$, $m$, $\mathcal{C}$. $N(\cp;d,c)$ $\forall \cp\in \mathcal{C}$, $P(\cp)$ $\forall \cp\in \mathcal{C}$. 
\STATE \textbf{Output}: $Q_p(\cp) \; \forall \cp\in \mathcal{C}$, Optionally $\braket{C(\vec{\gamma},\vec{\beta})}_h$
\STATE $Q_0(\cp) \gets 1\sqrt{2^n} \; \forall \cp \in \mathcal{C}$
\FOR{$\ell$ in $[1,p]$}
    \STATE $Q_{\ell}(\cp) \gets \sum_{d,c}  \cos\beta^{n-d}(-i\sin\beta)^d e^{-i\gamma c} Q_{\ell-1}(c) N(\cp;d,c) \quad \forall \cp \in \mathcal{C}$
\ENDFOR
\STATE If desired, $\widetilde{\braket{C}} \gets \sum_{\cp}  2^n P(\cp)|Q_{\ell}(\cp)|^2 \cp$
\\\hrulefill
\end{algorithmic}
\end{algorithm}

\emph{Time Complexity}: Given $N(\cp;d,c)$ as well as  $Q_{\ell-1}(\cp)$ for all $\cp$ in $\mathcal{C}$, the calculation of each amplitude $Q_{\ell}(\cp)$ using Eq.~\eqref{eq:homog_sop} is $\mathcal{O}(n|\mathcal{C}|)$. Thus, the calculation of all $|\mathcal{C}|$ amplitudes $Q_{\ell}(\cp)$ is $\mathcal{O}(n|\mathcal{C}|^2)$. Computing all $Q_{p}(\cp)$ elements then is $\mathcal{O}(n p |\mathcal{C}|^2)$. If desired, the evaluation of the cost function, given in Eq.~\eqref{eq:homog_sop_cost}, involves a sum over $\mathcal{C}$ elements, thus is $\mathcal{O}(|\mathcal{C}|)$.

Thus, if $|\mathcal{C}|$ is $O(\mathrm{poly}(n))$, then Algorithm \ref{alg:homog_apx_evol} is efficient. Indeed, we show such a case in the following section, demonstrating how to calculate the necessary pre-computations of $\mathcal{C}$, as well as $N(\cp;d,c)$ and $P(\cp)$ for all $\cp$ in $\mathcal{C}$.

\section{Cost distributions for Randomly Drawn CSPs}
\label{n_dists_random}

In this section we demonstrate how to compute $\mathcal{C}$ and $\widetilde{\braket{C}}$ for a number of random CSPs. For the Classical Homogeneous Proxy of Sec.~\ref{sum_of_paths_perspective}, we assumed that for all $\cp$ in $\mathcal{C}$, we are given distributions $N(\cp;d,c)$ that suitably estimate $n(x;d,c)$ for all $x$ with cost $\cp$. Obtaining these distributions can be viewed as a pre-computation step, and when derived from the properties of a particular problem class they may then be applied for any instance within. Here, we identify common random classes of optimization problems where $N(\cp;d,c)$ can be efficiently computed for the entire class. Particularly, we focus on CSPs with a fixed number~$m$ of Boolean clauses, each acting on $k$ variables selected at random from the set of $n$ variables. The types of allowed clauses is determined by the problem, for example SAT problems consider disjunctive clauses. We note that the analysis here generalizes a similar method in \cite{hogg00} applied to 3-SAT, allowing for easy extension to any CSP matching the above criteria. For these problems, the procedure is as follows, noting that \emph{all calculations done here are averaged over the entire class}: we first can rewrite the expected number of bitstrings with cost $c$ at distance $d$ from a random bitstring of cost $\cp$ as,
\begin{equation}\label{eq:rand_n_dist_start}
    N(\cp;d,c) = \binom{n}{d}P(c|d,\cp),
\end{equation}
where $P(c|d,\cp)$ represents the probability that a bitstring at distance $d$ from a bitstring with cost $\cp$ has cost $c$. This probability can then be rewritten as 
\begin{equation}\label{eq:p_of_cprime_given_d_and_c}
P(c|d,\cp) = \frac{P(\cp,c|d)}{P(\cp)} ,
\end{equation}
where $P(\cp,c|d)$ represents the probability that two bitstrings separated with Hamming distance $d$ have costs $c$ and $\cp$ and $ P(\cp)$ represents the probability that a randomly chosen bitstring has cost $\cp$. The numerator can be calculated as follows:
\begin{equation}
P(\cp,c|d) = \sum_{b=\max(0, \cp+c-m)}^{\min(\cp,c)} P(b, \cp-b, c-b | d),
\end{equation}
where $P(b, \cp-b, c-b | d)$ represents the probability that two randomly chosen bitstrings with Hamming distance $d$ have $b$ satisfied clauses in common, along with cost $\cp$ and $c$. This expression can be evaluated via the multinomial distribution
\begin{equation}
P(b, \cp-b, c-b | d) = \frac{m!}{b!(\cp-b)!(c-b)!(m+b-(\cp+c))!} \Pboth^b \Pone^{(\cp+c)-2b} \Pneither^{m+b-(\cp+c)},
\end{equation}
where $\Pboth$, $\Pone$, and $\Pneither$ represent the probability that a randomly selected clause is satisfied by both, one, or neither of the bitstrings separated by Hamming distance $d$, respectively. Since $\Pboth + 2\Pone + \Pneither = 1$, one only needs to calculate two of these three variables. 
The previous equations then allow computing $N(\cp;d,c)$ for any value $\cp$ as
\begin{equation}\label{eq:rand_n_dist_end}
N(\cp;d,c) = \binom{n}{d} \frac{1}{P(\cp)}\sum_{b=\max(0, \cp+c-m)}^{\min(\cp,c)} \frac{m!}{b!(\cp-b)!(c-b)!(m+b-(\cp+c))!} \Pboth^b \Pone^{(\cp+c)-2b} \Pneither^{m+b-(\cp+c)}.
\end{equation}

We summarize this approach in Algorithm~\ref{alg:homog_apx_precomp}.
\vspace{2mm}

\begin{algorithm}
\caption{Classical Homogeneous Proxy for QAOA Pre-computation for Randomly Drawn Cost Function}\label{alg:homog_apx_precomp}
\begin{algorithmic}
\STATE \textbf{Input}: $n$, $m$, Description of problem class 
\STATE \textbf{Output}: $\mathcal{C}$. $N(\cp;d,c)$, $P(\cp)$ $\forall \cp\in \mathcal{C}$
\STATE Determine a suitable $\mathcal{C}$ via Sec.~\ref{cost_distributions}.
\STATE Compute $\Pboth$, $\Pone$, $\Pneither$ and $P(\cp) \; \forall \cp\in \mathcal{C}$ given the problem class
\STATE Use these values to compute $N(\cp;d,c) \; \forall \cp\in \mathcal{C}$ via Eq.~\eqref{eq:rand_n_dist_end}
\\\hrulefill
\end{algorithmic}
\end{algorithm}

\emph{Time Complexity}: There are $|\mathcal{C}|$ distributions $N(\cp;d,c)$ with $(n+1)|\mathcal{C}|$ elements each. For fixed $\cp$, $d$, and $c$, we must sum over $|\mathcal{C}|$ terms that can be evaluated in $\mathcal{O}(|\mathcal{C}|)$ steps, given by the time complexity of evaluating factorials of  $\mathcal{O}(|\mathcal{C}|)$. Thus the evaluation of all distributions is $\mathcal{O}(n|\mathcal{C}|^4)$. We once again note that if $|\mathcal{C}|$ is $O(\mathrm{poly}(n))$, Algorithm \ref{alg:homog_apx_precomp} runs in polynomial time.

We now demonstrate Algorithm~\ref{alg:homog_apx_precomp} for several common problem classes.

\subsection{MaxCut}\label{example_maxcut}

We first analyze MaxCut on Erd\H{o}s-R\'{e}nyi random graphs~$\mathcal{G}(n,p_e)$. In this  model, a graph $G$ is generated by scanning over all possible $\binom{n}{2}$ edges in an $n$ node graph, and including each edge with independent probability $p_e$. The MaxCut problem on $G$ is to partition the $n$ nodes into two sets such that the number of cut edges crossing the partition is maximized. With respect to the class~$\mathcal{G}(n,p_e)$ of graphs the cost function to be maximized over configurations $x$ is  
\begin{equation}\label{eq:maxcut_cost_fn}
    c(x) = \sum_{\langle i,j \rangle \in \binom{n}{2}} e_{ij} x_i \oplus x_j,
\end{equation}
where $e_{ij}$ are independent binary random variables that take on value $0$ with probability $1-p_e$ and $1$ with probability $p_e$. We use this cost function to evaluate the relevant distributions in Eq.~\eqref{eq:rand_n_dist_end}. We start by noting for a fixed graph $G$ with $m$ edges we have $0\leq c(x) \leq m$, and the expected number of edges $\mathbb{E}[m]$ in graphs drawn from~$\mathcal{G}(n,p_e)$ is $p_e \binom{n}{2}$, with a standard deviation proportional to to the square root of this value as determined by the binomial distribution. Hence, as $n$ becomes large the expected number of edges concentrates to $p_e \binom{n}{2}$, and so for simplicity in the remainder of this work we set $\mathcal{C}=\{0,1,\dots,\lceil \mathbb{E}[m]\rceil \}$. Note that in practice, to accommodate the possibility of a given instance with maximum cut greater than this quantity, one may 
increase 
$\mathcal{C}$ by several standard deviations as desired.

We can then easily calculate $P(\cp)$, the probability a random bitstring has cost $\cp$ for Eq.~\eqref{eq:maxcut_cost_fn}. For a bitstring drawn uniformly at random, the probability of satisfying any given clause $x_i \oplus x_j$ is $1/2$, as there are two satisfying assignments ($01$ and $10$), and two non-satisfying assignments ($00$ and $11$). Thus, the probability $P(\cp)$ also follows a binomial distribution and is simply
\begin{equation}\label{eq:pc_maxcut}
    P(\cp) = \left(\frac{1}{2}\right)^m \binom{m}{\cp}.
\end{equation}
We can then calculate $\Pboth$, $\Pone$ and $\Pneither$, where we show all three for didactic purposes (since $\Pboth + 2\Pone + \Pneither =1$). To see this, consider two randomly chosen bitstrings $y$ and $z$, separated by Hamming distance $d$. We note that there are $(n-d)$ bits in common between $y$ and $z$ and $d$ bits different. Thus $y_i \oplus y_j = z_i \oplus z_j$ requires that either both $i$ and $j$ are from the set of $(n-d)$ same bits or the set of $d$ different bits, which has probability
\begin{equation}
    \Psame = \frac{\binom{n-d}{2}}{\binom{n}{2}} + \frac{\binom{d}{2}}{\binom{n}{2}}.
\end{equation}
From this expression, we can easily see that $\Pboth=\Pneither=\frac{1}{2}\Psame$. Since $\Psame$ represents the probability that $y_i \oplus y_j = z_i \oplus z_j$, for a random clause and a random bitstring, the probability that $y_i \oplus y_j=1$ is $1/2$ and the same is true for $y_i \oplus y_j=0$ by symmetry. Thus we have
\begin{equation}
    \Pboth =  \Pneither = \frac{1}{2} \left( \frac{\binom{n-d}{2}}{\binom{n}{2}} + \frac{\binom{d}{2}}{\binom{n}{2}} \right).
\end{equation}
For completion, we can calculate $\Pone$, which is $(1-\Psame-\Pboth)/2$ as shown in Sec.~\ref{sum_of_paths_perspective}. For this term we need $y_i \oplus y_j \neq z_i \oplus z_j$, which can be accomplished if one of $i$ or $j$ is chosen from the $(n-d)$ bits in common and the other is chosen from the $d$ differing bits. This probability is 
\begin{equation}
    \Pdif = \frac{\binom{n-d}{1}\binom{d}{1}}{\binom{n}{2}}.
\end{equation}
Thus, $\Pone$, which is the probability of specifically $y$ satisfying and $z$ not satisfying (or vice versa) is half of $\Pdif$, so
\begin{equation}
    \Pone = \frac{1}{2} \left( \frac{\binom{n-d}{1}\binom{d}{1}}{\binom{n}{2}} \right).
\end{equation}

Using these quantities $N(\cp;d,c)$ is then obtained from Eq.~\ref{eq:rand_n_dist_end}.

\subsection{MaxE3Lin2/Max-3-XOR}
We next consider the MaxE3Lin2 problem 
which generalizes MaxCut. QAOA for MaxE3Lin2 was analyzed by Farhi et al. \cite{farhi14}, who showed an advantage over classical approximation algorithms, only to inspire better classical approaches \cite{barak2015beating,hastings19}. The analogous random class of MaxE3Lin2 problems has cost function
\begin{equation}\label{eq:maxe3lin2}
    c(x) = \sum_{a<b<c} d_{abc}(x_a \oplus x_b \oplus x_c \oplus z_{abc}),
\end{equation}
where the $z_{abc}$, $,d_{abc}$ are independent random variables with equal probability of being $0$ or $1$. 

Using Eq.~\eqref{eq:maxe3lin2} we can again calculate the relevant probability distributions. First note that a random bitstring $x$ will satisfy a individual clause (i.e. term in the sum) with probability $1/2$, as $(x_a + x_b + x_c) \mod 2$ has an equal probability to be $0$ or $1$. Thus,
\begin{equation}
    P(\cp) = \left(\frac{1}{2}\right)^m \binom{m}{\cp},
\end{equation}
using the binomial distribution. Then we note that, as in Sec.~\ref{example_maxcut}, the probability of $(y_a + y_b + y_c) \mod 2 = (z_a + z_b + z_c) \mod 2$ for two random bitstrings with $d(x,y)=d$ is given by 
\begin{equation}
    \Psame = \frac{\binom{n-d}{3}+\binom{d}{2}\binom{n-d}{1}}{\binom{n}{3}},
\end{equation}
since the value of $(x_a + x_b + x_c) \mod 2$ is preserved if none or two of $x_a,x_b,x_c$ are flipped, which is satisfied if $a,b,c$ are all from the $n-d$ identical bits, or two of $a,b,c$ are chosen from the $d$ differing bits. Thus, we can easily compute $\Pboth=\Pneither=\Psame/2$, since there is an equal chance $(y_a + y_b + y_c) \mod 2 = (z_a + z_b + z_c) \mod 2 = 0/1$. $\Pone$ can then be calculated by a similar argument or by taking $\Pone=(1-\Pboth-\Pneither)/2$. We note that the probability distributions calculated here are exactly equivalent to those for the Max-3-XOR problem, where all $z_{abc}$ in Eq.~\eqref{eq:maxe3lin2} are taken to be $1$.

\subsection{MaxEkLin2/Max-k-XOR}

The MaxEkLin2 problem is a further generalization for each fixed for fixed $k$ of MaxE3Lin2, where we replace the $a,b,c$ with $a_1,...,a_k$ in the cost function and the sum is taken over hyperedges of size~$k$. This class of problems was previously studied for QAOA in~\cite{marwaha2022bounds,chou2022limitations}. For each $k$, $P(\cp)$ is the same as for MaxE3Lin2 above. However, $\Psame$ is given by
\begin{equation}
    \Psame = \frac{\sum_{h=0}^{\lfloor k/2 \rfloor}\binom{d}{2l}\binom{n-d}{k-2l}}{\binom{n}{k}},
\end{equation}
where the terms in the sum represent all possible ways to choose an even number of bits to flip from the $k$ bits in the clause out of $d$ total bits. Then, again, we have again $\Pboth=\Pneither=\Psame/2$ and $\Pone=(1-\Pboth-\Pneither)/2$. A similar calculation for the Max-k-XOR problem again yields identical probability distributions.

\subsection{Rand-k-SAT}

The case of random k-SAT is analyzed by Hogg in \cite{hogg00}. This cost function is defined as the sum over $m$ clauses of a logical OR of $k$ variables randomly drawn from a set of $n$ variables, each of which may be negated. 
In the notation used in this paper, the distributions of interest are
\begin{equation}
    P(\cp) = 2^{-km}(2^{k}-1)^{m-\cp}\binom{m}{\cp},\quad \Pboth=\frac{2^{-k}\binom{n-d}{k}}{\binom{n}{k}},\quad \Pone=2^{-k}-\Pboth,\quad
    \Pneither=1-2\Pone-\Pboth.
\end{equation}

\section{Homogeneous Heuristic for Parameter Setting}\label{homogeneous_parameter_setting}

Leveraging the Classical Homogeneous Proxy for QAOA, here we propose a strategy for finding good algorithm parameters, which we call the Homogeneous Heuristic for Parameter Setting, as pictured in Fig.~\ref{fig:flowchart} and formalized in Algorithm~\ref{alg:homog_param_set}. 

\vspace{2mm}
\begin{algorithm}[h]
\caption{Homogeneous Heuristic for Parameter Setting}\label{alg:homog_param_set}
\begin{algorithmic}
\STATE \textbf{Input}: $p$, $\vec{\gamma_{\mathrm{in}}}$, $\vec{\beta_{\mathrm{in}}}$, $n$, $m$, $\mathcal{C}$. $N(\cp;d,c)$ $\forall \cp\in \mathcal{C}$, $P(\cp)$ $\forall \cp\in \mathcal{C}$, constraints on $(\vec{\gamma}, \vec{\beta})$. 
\STATE \textbf{Output}: $\vec{\gamma_{\mathrm{out}}}$, $\vec{\beta_{\mathrm{out}}}$
\STATE{$\vec{\gamma}, \vec{\beta} \gets \vec{\gamma_{\mathrm{in}}}, \vec{\beta_{\mathrm{in}}}$ }
\WHILE{Desired}
    \STATE{1) Using any suitable parameter update scheme, perform one update of $(\vec{\gamma}, \vec{\beta})$}
    \STATE{2) Evaluate the homogeneous parameter objective function (Eq.~\eqref{eq:homog_sop_cost}) using the Classical Homogeneous Proxy for QAOA for all $(\vec{\gamma}, \vec{\beta})$ required to perform next parameter update in 1)}
\ENDWHILE
\STATE{$\vec{\gamma_{\mathrm{out}}}, \vec{\beta_{\mathrm{out}}} \gets \vec{\gamma}, \vec{\beta}$ }
\\\hrulefill
\end{algorithmic}
\end{algorithm}

Here, a ``suitable parameter update scheme'' is intended to encapsulate a wide variety of general or specific approaches proposed for this in the literature, ``Desired'' denotes that the while loop can be iterated until the update scheme terminates or some desired convergence criteria is met, and ``Constraints on $(\vec{\gamma}, \vec{\beta})$'' denotes any restrictions on the domain of values allowed for $\vec{\gamma}, \vec{\beta}$, including restrictions to schedules of a prescribed form such as linear ramps introduced in Sec.~\ref{numerical_overlaps}. 

With the heuristic, we replace the typical cost expectation value with the homogeneous parameter objective function, where each function evaluation takes time as determined by Algorithms~\ref{alg:homog_apx_evol} and~\ref{alg:homog_apx_precomp}. This heuristic is purposefully defined in broad terms, in order to maintain complete freedom in the choice of parameter update schemes. Thus, one can still apply a myriad of approaches explored in parameter setting literature, such as parameter initialization, re-parameterization, and the use of different global or local optimization algorithms. 

On the other hand, we emphasize that while our approach can significantly speed up the parameter setting task, it is by no means a panacea. Indeed, in cases where the number of parameters to optimize grows with the problem size (e.g., when $p=n$), this problem suffers generically from the curse of dimensionality, as well as other potential difficulties such as barren plateaus or plentiful local optima. Hence the incorporation of a variety of parameter setting strategies or approximations that seek to ameliorate these difficulties within our approach is well-motivated.

\section{Numerically Investigating the Classical Homogeneous Proxy for QAOA for MaxCut}\label{validity_of_approximation}

In this section, we explore the application of the Classical Homogeneous Proxy for QAOA to MaxCut on Erd\H{o}s-R\'{e}nyi $\mathcal{G}(n, p_e)$ model graphs as considered in Sec.~\ref{example_maxcut}. We first numerically study the accuracy of replacing $n(x;d,c)$ distributions with $N(\cp;d,c)$ distributions as calculated via the methods presented in Sec.~\ref{n_dists_random}. We then numerically show that the proxy maintains large overlaps with full classical statevector simulation of QAOA for certain parameter schedules. Finally, we provide a toy example for a small graph at $p=3$, empirically showing that the homogeneous and typical parameter objective function correlate significantly for important parameter regimes.

\subsection{Viability of Replacement Distance and Cost Distributions}

\begin{figure}[!htb]
    \centering
    \includegraphics[width=0.55\textwidth]{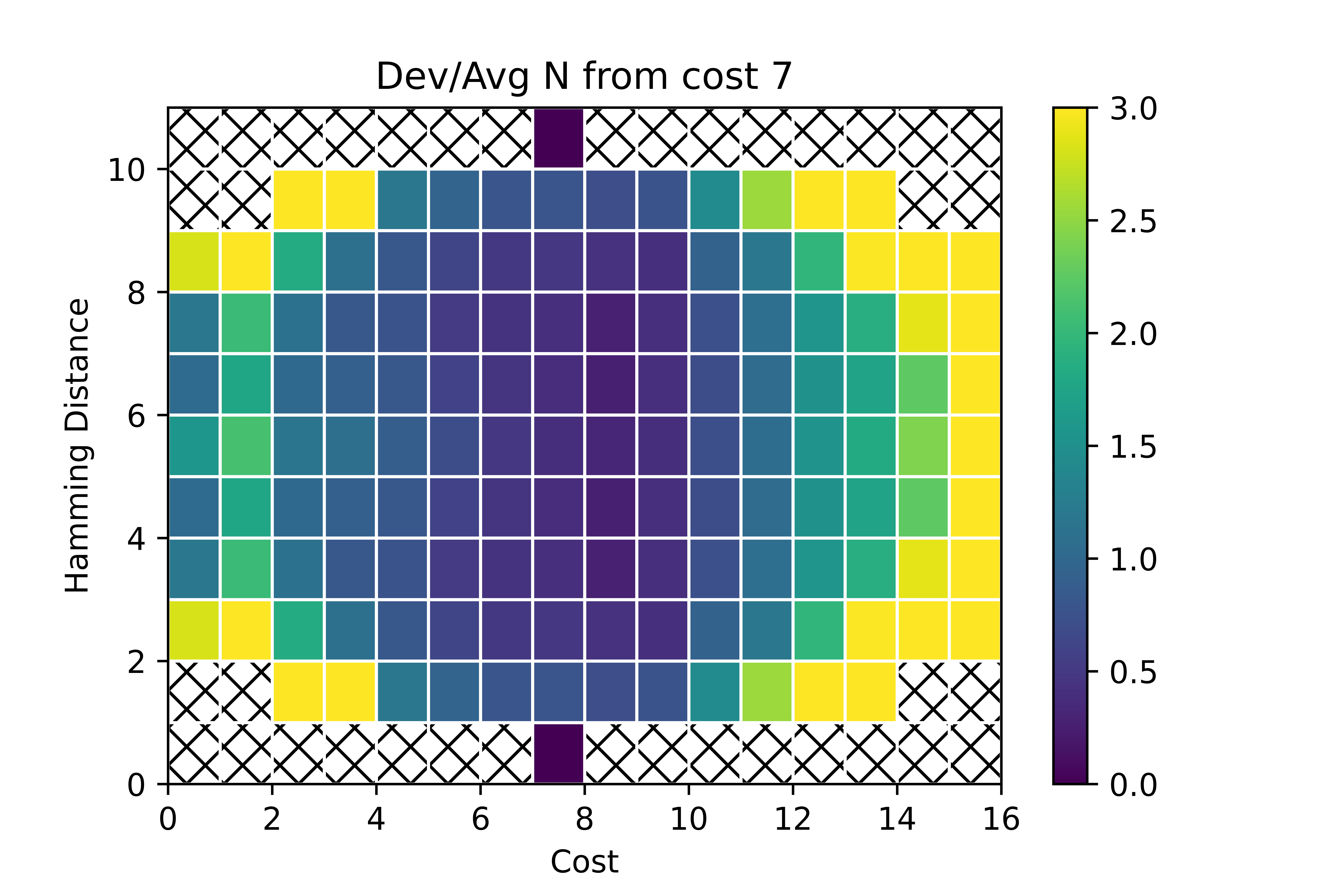}
    \caption{Heat map of the standard deviation/mean of $N(\cp;d,c)$ distributions for 10 instances of $\mathcal{G}(10,1/3)$ graphs, with $\cp$ fixed at $7$. Hatched out squares correspond to those elements of the distribution for which $N(\cp;d,c)$ is always $0$, meaning that no bitstring at distance $d$ from a bitstring with cost $7$ has cost $c$. $\cp=7$ was chosen because for $\mathcal{G}(10,1/3)$ graphs, $P(\cp)$ is peaked near $7$, so this represents a typical instance of an $N(\cp;d,c)$ distribution.}
    \label{fig:n10_er_p3overn_n_dev_ten_graphs}
\end{figure}

For these experiments, we first choose ten $\mathcal{G}(10,1/3)$ Erd\H{o}s-R\'{e}nyi model graphs, and calculate the $n(x;d,c)$ for all bitstrings $x$ in $\{0,1\}^n$, all d in $[0,n]$ and all $c$ in $[0,m]$, with $m$ being an upper bound on the maximum cost (here we use $p_e \binom{n}{2}$, as described in Sec.~\ref{example_maxcut}, in this case $m=15$). For each cost $\cp$, we consider $n(x;d,c)$ for all states $x$ with cost $\cp$ across all $10$ graphs. In order to evaluate the viability of replacing $n(x;d,c)$ with $N(\cp;d,c)$, we present the following intuition: the better that $N(\cp;d,c)$ estimates the average of $n(x;d,c)$ over all $x$ with $c(x)=\cp$, and the less that $n(x;d,c)$ deviates over $x$ with $c(x)=\cp$, the better $N(\cp;d,c)$ should estimate $n(x;d,c)$ for all $x$ with $c(x)=\cp$. We thus aim to numerically demonstrate the extent to which both the analytically derived $N(\cp;d,c)$ estimate the average $n(x;d,c)$ and to which $n(x;d,c)$ deviates from its average. We first demonstrate the latter. To do this, we take the element-wise averages of these distributions. This average is one way of computing the distributions $N(\cp;d,c)$, as described in Sec.~\ref{sum_of_paths_perspective}. We also take the element-wise standard deviations of these distributions. In Fig.~\ref{fig:n10_er_p3overn_n_dev_ten_graphs} we display the element-wise ratio of standard deviation to mean of these distributions for $\cp=7$, chosen because $P(\cp)$ is maximal near $7$, such that this term has large weight in Eq.~\eqref{eq:homog_sop}.

From the figure, we see that for costs near $m/2$ ($7.5$) and distances near $n/2$ ($5$), there is minimal deviation in the $N(\cp;d,c)$ distributions among multiple instances of Erd\H{o}s-R\'{e}nyi model graphs and multiple bitstrings $x$ with cost $\cp$. We note that the relative deviation is highest at the edges of the plot, where $d \rightarrow 0$,  $d \rightarrow n$, $c \rightarrow c_{min}$, and $c \rightarrow c_{max}$. We note, however, that at these points, the expected value of $n(x;d,c)$ is smaller, such that the contribution of these deviations to the sum in Eq.~\eqref{eq:pre_homog_sop} is less than those with distance and cost near the center of the distribution. As an example, there are $\binom{10}{5}=252$ bitstrings at distance $5$ from a given bitstring $x$, as opposed to $\binom{10}{1}=10$ bitstrings at distance $1$. A similar argument can be made using the values of $P(\cp)$ derived in Sec.~\ref{n_dists_random}. This numerical evidence thus suggests that replacing $n(x;d,c)$ with $N(\cp;d,c)$ determined via an averaging over all $x$ with cost $\cp$ over multiple instances may introduce deviations from Eq.~\eqref{eq:homog_sop} precisely in cases that contribute to less Eq.~\eqref{eq:general_amp_sop}, allowing for near-homogeneous evolution.

\begin{figure}[!htb]
    \centering
    \includegraphics[width=0.92\textwidth]{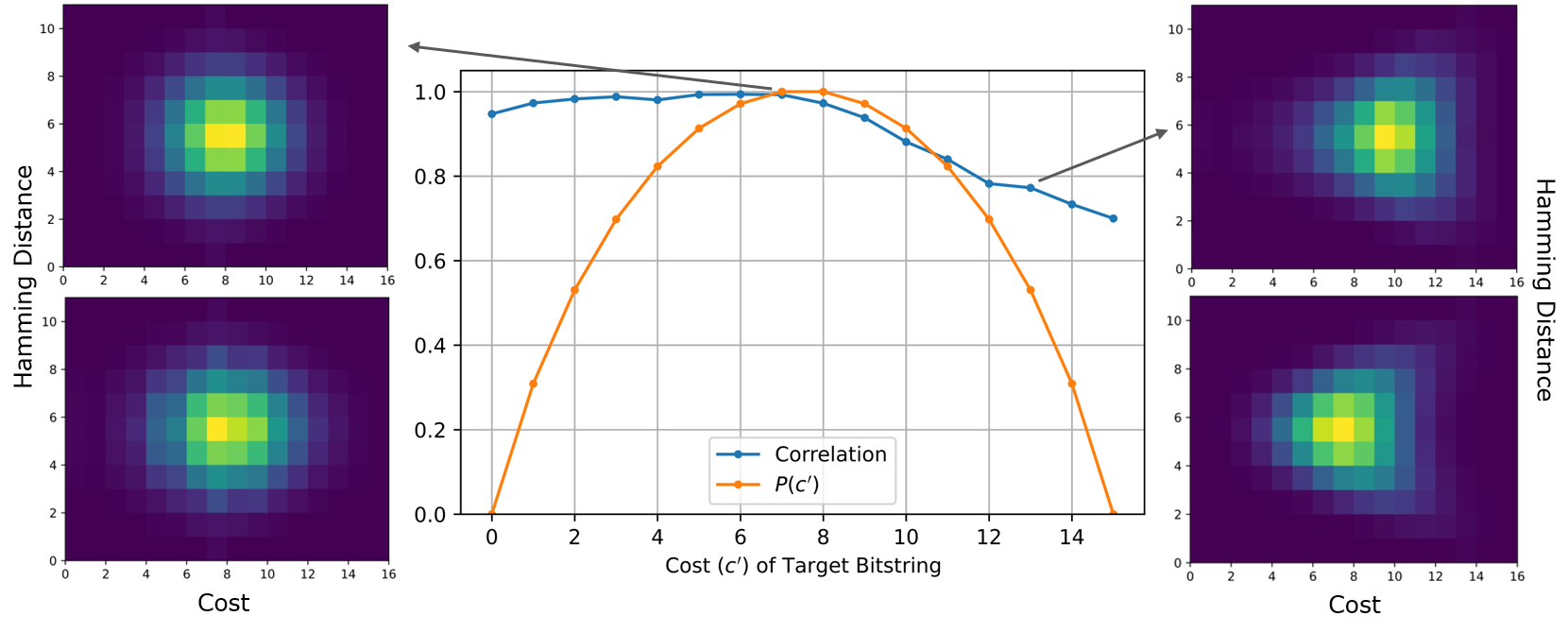}
    \caption{(middle) Pearson Correlation Coefficients between $N(\cp;d,c)$, calculated through the averaging over 10 $\mathcal{G}(10,1/3)$ graph instances and through the analytical method presented in Sec.~\ref{n_dists_random}. $P(\cp)$, the probability of a random bitstring having cost $\cp$ is also displayed to elucidate dominant terms in the sum of Eq.~\eqref{eq:homog_sop}. Inserts display the distribution for fixed $\cp$ for (left) $\cp=7$ using (top) the analytical approach and (bottom) the averaging approach. (right) insert displays the same for $\cp=13$.}
    \label{fig:n_apx_comparison}
\end{figure}

This result provides evidence that the deviation in $n(x;d,c)$ is small for the \lq\lq bulk\rq\rq\ of contributions to the sum in Eq.~\eqref{eq:general_amp_sop}, such that we can replace $n(x;d,c)$ with $N(\cp;d,c)$, the average over $x$ with $c(x)=\cp$ for the \emph{entire class} of problems. We then would like to understand how the analytically derived expressions for $N(\cp;d,c)$ in  Sec.~\ref{n_dists_random} estimate these averaged distributions. We perform the comparison of $N(\cp;d,c)$ computed by averaging and the methods in Sec.~\ref{n_dists_random} for MaxCut on Erd\H{o}s-R\'{e}nyi model graphs in Fig.~\ref{fig:n_apx_comparison}, showing the Pearson correlation coefficient between the two distributions for each $\cp$ in $[0,m]$. We likewise display $P(\cp)$, in order to elucidate the dominant distributions in the sum of Eq.~\eqref{eq:homog_sop}. 

From the figure we see that for dominant terms (with high $P(\cp)$), the two distributions align well visually, corresponding to a high correlation coefficient. For less important terms, the analytical distributions do not match the average over many instances, but the effect of this mismatch may be reduced due to the lesser weight on these terms as determined by $P(\cp)$ in the sums of equation Eq.~\eqref{eq:homog_sop}.

Combined, Figs.~\ref{fig:n10_er_p3overn_n_dev_ten_graphs} and~\ref{fig:n_apx_comparison} show that for dominant terms, there is little deviation in $n(x;d,c)$ distributions from their average $N(\cp;d,c)$, and the analytically computed distributions match these average distributions well. Thus, they together indicate that the analytical methods for calculating $N(\cp;d,c)$ should approximate $n(x;d,c)$ with $c(x)=\cp$ for terms that dominate the sum in Eq.~\eqref{eq:homog_sop}.

\subsection{Numerical Overlaps}\label{numerical_overlaps}

\begin{figure}[!htb]
    \centering
    \includegraphics[width=0.65\textwidth]{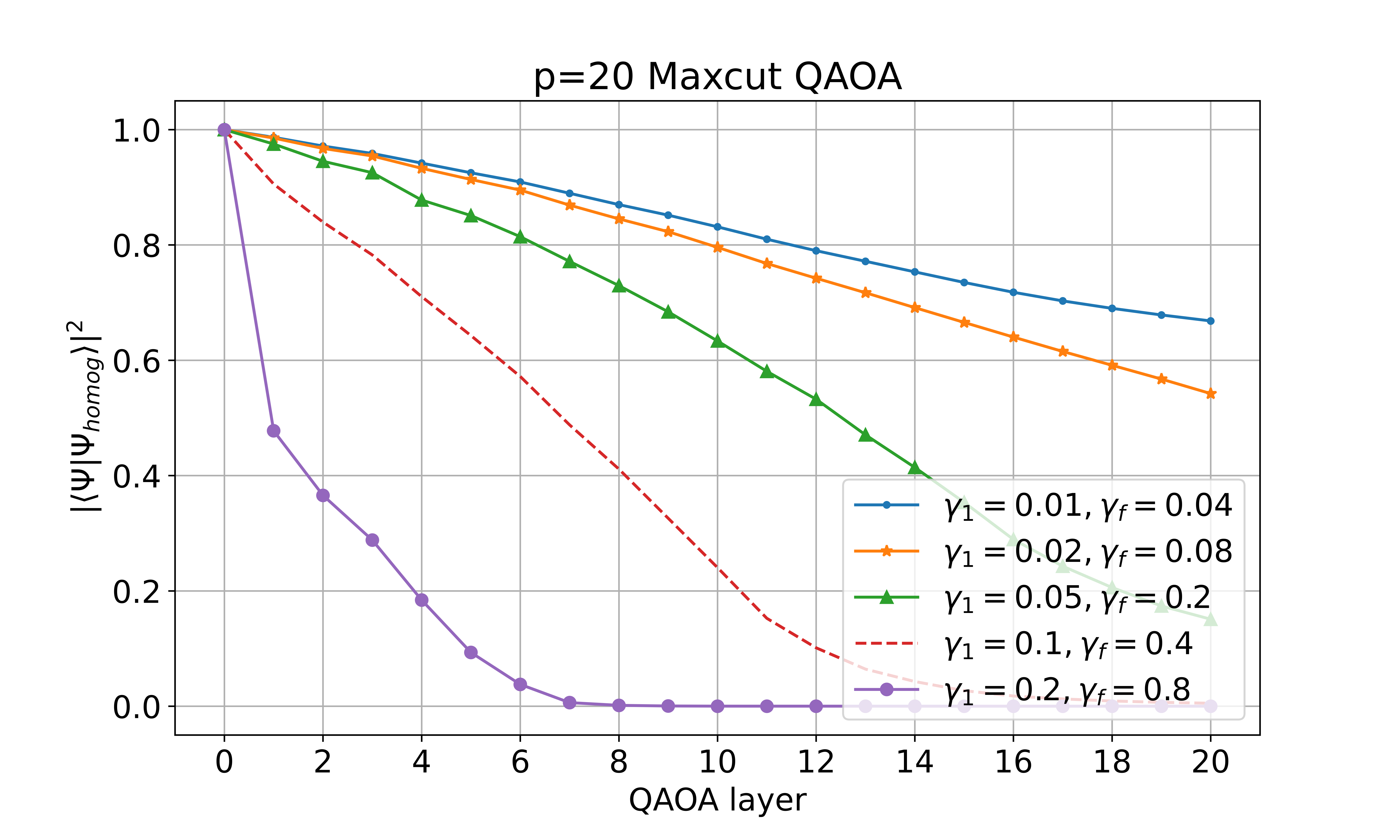}
    \caption{Squared overlap between $p=20$ MaxCut QAOA run on a $\mathcal{G}(8,1/2)$ Erd\H{o}s-R\'{e}nyi model graph instance using full state simulation vs. Eq.~\eqref{eq:homog_sop}, with distributions calculated as in Sec.~\ref{n_dists_random} as a function of current QAOA layer. A fixed linear ramp parameter schedule is chosen, with $\gamma$ increasing and $\beta$ decreasing at each step. Each curve corresponds to differing values of $\gamma_1$ and $\gamma_f$}
    \label{fig:ramp_homog}
\end{figure}

To further investigate the viability of the Classical Homogeneous Proxy for QAOA, we perform numerical simulations of the proxy on MaxCut on Erd\H{o}s-R\'{e}nyi graphs $\mathcal{G}(10, .5)$, and display the squared overlap between classical full statevector simulation and the proxy as a function of $p$ for various parameter schedules motivated by QAOA literature.
For this analysis, we choose linear ramp parameter schedules, inspired by quantum annealing. In particular, we fix a starting and ending point for $\vec{\gamma}$ and $\vec{\beta}$, which is kept constant regardless of $p$, and the schedule is defined as 
\begin{equation}\label{eq:ramp_schedule}
    \gamma_{\ell} = \gamma_1 + (\gamma_f - \gamma_1)\frac{\ell}{p}, \quad
    \beta_{\ell} = \beta_1 + (\beta_f - \beta_1)\frac{\ell}{p} 
\end{equation}
for each layer ${\ell}$. Given these linear ramp schedules, the squared overlaps between the replace and original quantities are calculated as follows. First,  statevector simulation was performed using HybridQ, an open-source package for large-scale quantum circuit simulation \cite{mandra21}. Next, the $N(\cp;d,c)$ distributions are computed according to Eq.~\ref{n_dists_random}. Then, for the purposes of comparison, we compute the proxy slightly differently from above, by starting with the uniform superposition over all $2^n$ bitstrings and simply plugging in $N(c(x);d,c)$ for all $n(x;d,c)$ in Eq.~\eqref{eq:pre_homog_sop}, keeping all $2^n$ amplitudes $q^{\ell}(x)$ at each step $\ell$. This allows us to compute the overlap between the proxy and true state using standard quantum state overlap, although we lose the gain in complexity from performing the proxy on only the set of unique costs, as we are working with $2^n$ bitstrings rather than $|\mathcal{C}|$ costs. Then, using this method, we plot the squared overlaps between the replace and original quantities as a function of $p$ with varying values of $\gamma_1$ and $\gamma_f$ in Fig.~\ref{fig:ramp_homog}.

From the figure, we can see that the overlap gradually decreases as the number of QAOA layers increases. However, the decline is less dramatic when $\gamma_1$ and $\gamma_f$ are lower in magnitude. Thus, we see that as we move towards the $\gamma \ll 1$ regime for these problems (or, more precisely $\gamma \ll \|C\|$~\cite{hadfield2021analytical}), the true QAOA state remains closer to the proxy even as the algorithm progresses, as remarked in Sec~\ref{homogeneous_approximation}. We stress that this behavior is empirical and the numerics are limited to the MaxCut examples analyzed presently. This behavior does however align with the analytical fact mentioned in Sec.~\ref{homogeneous_approximation} that to first order in $\gamma$, QAOA states are Perfectly Homogeneous for strictly k-local Hamiltonians.

\subsection{Parameter Objective Function Landscapes at Low Depth}\label{landscapes}

\begin{figure}[!htb]
    \centering
    \includegraphics[width=0.48\textwidth]{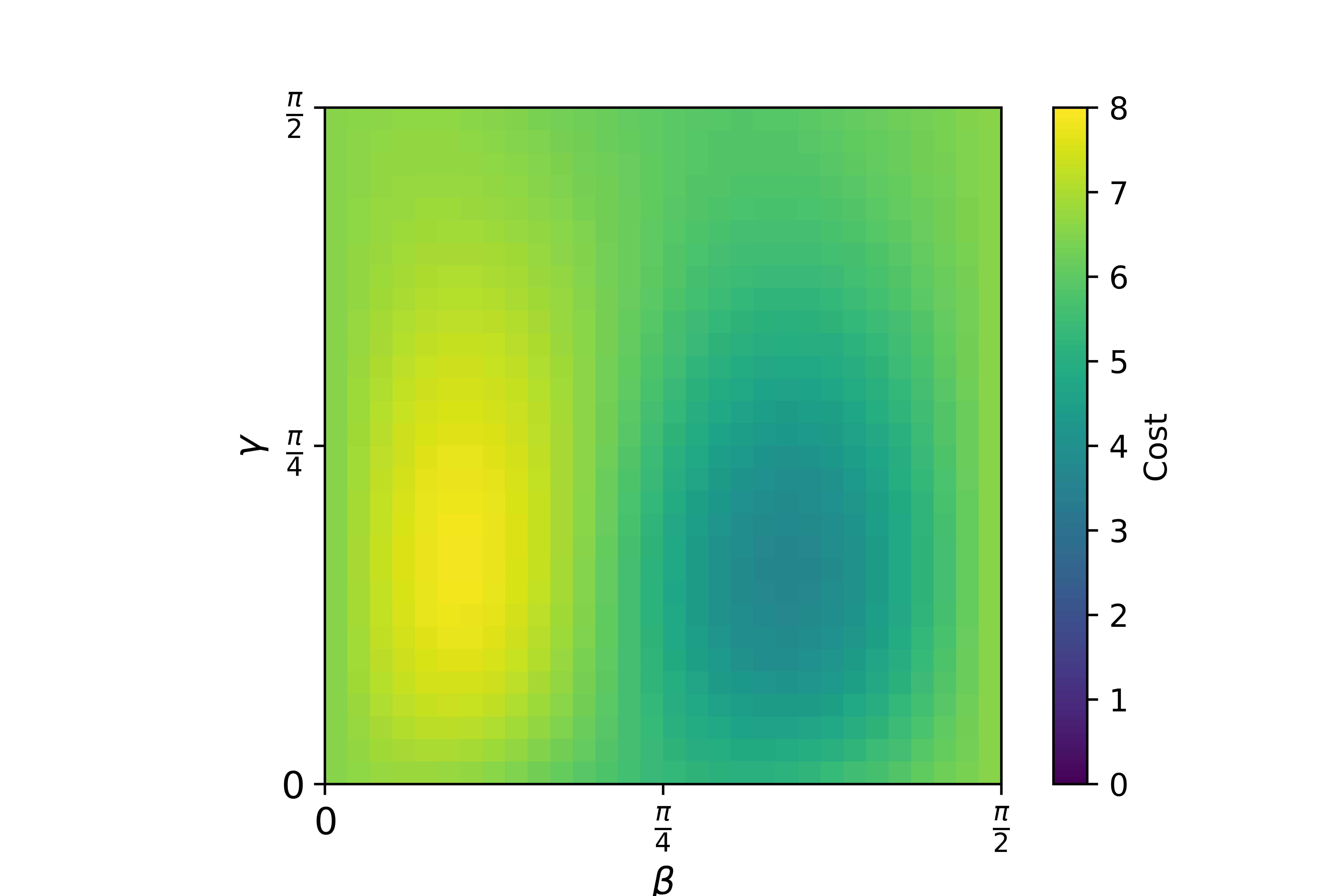}
    \includegraphics[width=0.48\textwidth]{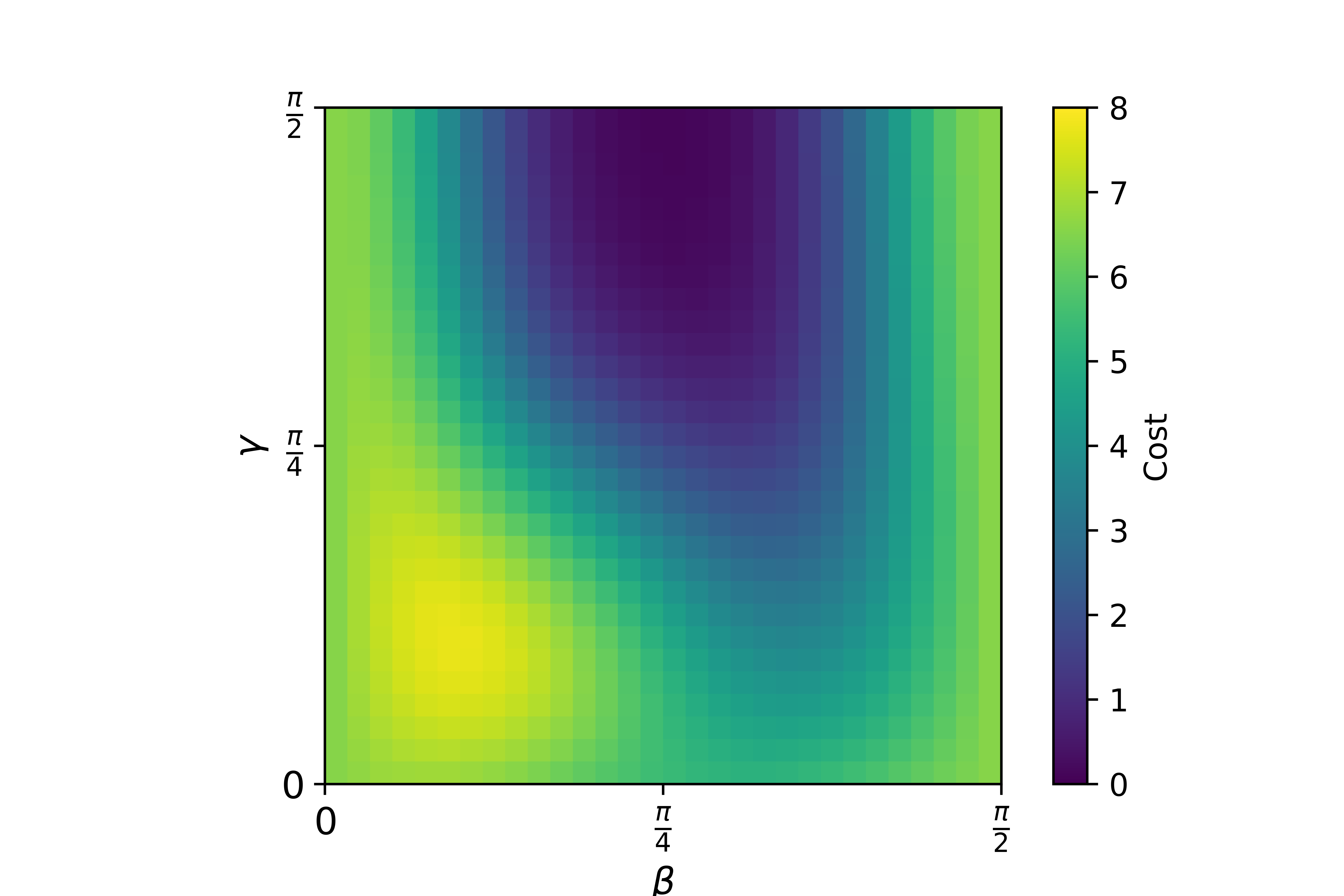}
    \caption{Parameter objective function landscapes displayed as heat maps as a function of $\gamma_3$ and $\beta_3$ for $p=3$ MaxCut QAOA on $\mathcal{G}(8,1/2)$ Erd\H{o}s-R\'{e}nyi model graph with $\gamma_1$, $\gamma_2$, $\beta_1$ and $\beta_2$ fixed. (left) The typical parameter objective function, computed via classical full statevector simulation. (right) The homogeneous parameter objective function computed via the Classical Homogeneous Proxy for QAOA.}
    \label{fig:n8_er_landscapes}
\end{figure}

In order to provide an explicit illustrative example for the efficacy of the Homogeneous Heuristic for Parameter Setting in Alg.~\ref{alg:homog_param_set}, we depict the both the typical and homogeneous parameter objective function as a function of $\gamma_3$ and $\beta_3$ for a randomly drawn $\mathcal{G}(8,1/2)$ graph and QAOA with $p=3$. In this example, our aim is to visualize similarities between the two parameter objective functions rather than to exhaustively find optimal parameters for the graph. As such, we borrow $\gamma_1$, $\gamma_2$, $\beta_1$ and $\beta_2$, optimized from a $20$ node instance in Sec.~\ref{low_depth}. These landscapes are shown in Fig.~\ref{fig:n8_er_landscapes}. 

It is visually clear that the two landscapes have significant differences. For the typical parameter objective function, there exists a clearly defined, Gaussian-like peak (yellow) and valley (blue). For the homogeneous parameter objective function, there exists a similarly-located peak, albeit vertically compressed, and the corresponding valley comprises almost the entire rest of the landscape. However, we can see that in the small $\gamma$ regime in particular, the landscapes qualitatively look very similar. This behavior is suggested by Fig.~\ref{fig:ramp_homog}, where we see quantitatively that the extent to which the Classical Homogeneous Proxy for QAOA overlaps with the true QAOA evolution grows as $\gamma$ decreases. 
Additionally, as seen in previous studies of QAOA parameters \cite{zhou20, crooks18, shaydulin21_qaoakit, lotshaw21}, optimal values of $\gamma$ tend to remain relatively small (exact values are not described as they depend on $n$, $p$, and the cost function being used), especially at the beginning of the algorithm. This suggests that, while the homogeneous and typical parameter objective functions may deviate significantly in general, they maintain significant correlations in \emph{relevant} parameter regimes, specifically those which are near the maximum in the landscape. Indeed, for the task of parameter setting, we expect qualitative feature mapping of the landscape to be much more important than a precise matching of objective function values.

It is also worthwhile to consider the difference in computational resources needed to produce the two plots. For the typical parameter objective function, in order to classically evaluate the evolution of the algorithm
in each cell of the presented 30 by 30 landscapes, we perform full statevector simulation. Farhi et al. show in \cite{farhi14}, that in order to compute Pauli observable expectation values for the typical parameter objective function, one only needs to include qubits that are within the reverse causal cone generated by the qubits involved in the observable and the quantum circuit implementing QAOA. However, for the example analyzed here, at $p=3$, this reverse causal cone includes all $8$ qubits for each observable, so in order to classically compute the evolution, we perform a full-state simulation. Thus, deriving evolution of the proxy took roughly one-fiftieth of the time required for simulating full QAOA. We note that it is possible to efficiently evaluate each cell on an actual quantum computer, and that if one only wants expectation values given parameters rather than the full state evolution, there are more efficient classical methods (e.g. \cite{farhi14,wang18}). Current difficulties for this approach, however include noise resulting both from finite sampling error as well as the effects of imperfect quantum hardware.

\section{Results}\label{results}

In this section we present numerical evidence supporting the Homogeneous Heuristic for Parameter Setting, again using MaxCut on Erd\H{o}s-R\'{e}nyi model graphs as a target application. Due to the array of possible techniques implementing the parameter update scheme as mentioned in Sec.~\ref{homogeneous_parameter_setting}, we do not wish to provide an exhaustive comparison of the heuristic to previous literature, but rather demonstrate regimes where the heuristic provides parameters that are either comparable with previous results, or that yield increasing performance out to larger values of $p$ where we are not aware of prior methods in the literature successfully returning well-performing parameter schedules.

\subsection{Global Optimization at Low-Depth}\label{low_depth}
 
\begin{figure}[!htb]
    \centering
    \includegraphics[width=0.85\textwidth]{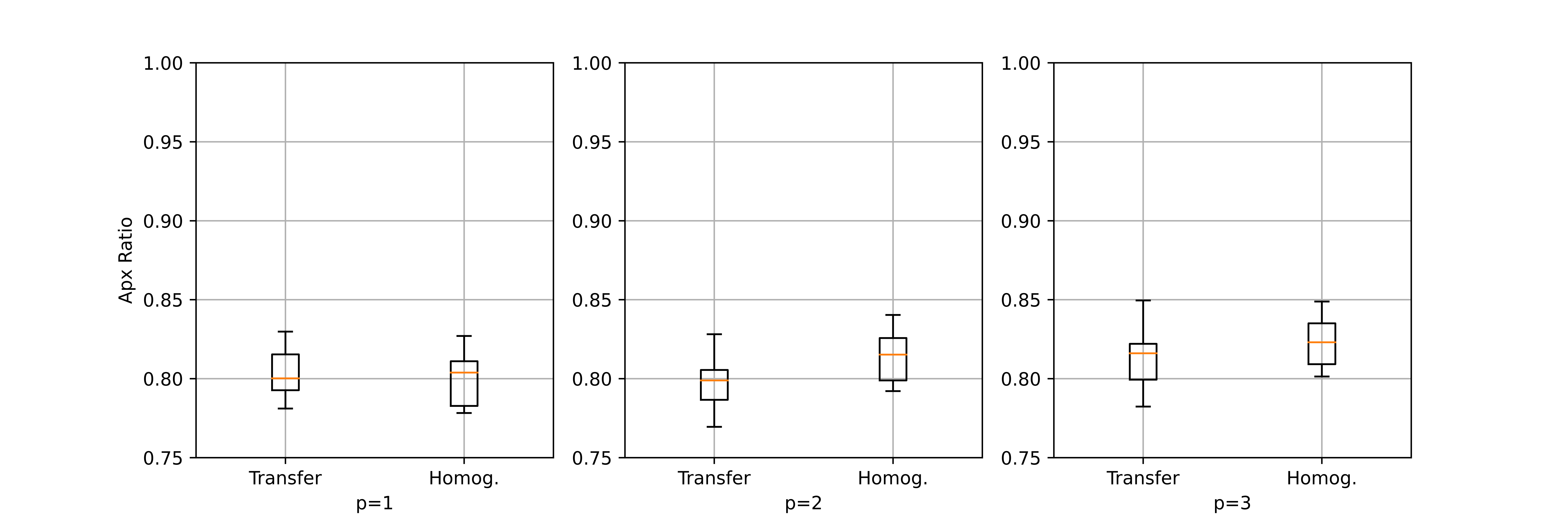}
    \caption{Box plots of approximation ratios obtained by the transfer of median of optimal parameters from $10$ $\mathcal{G}(9,1/2)$ graphs to $10$ $\mathcal{G}(20,1/2)$ graphs vs parameters found via homogeneous parameter setting for the same $10$ $\mathcal{G}(20,1/2)$ graphs, for $p=$ $1$,$2$, and $3$.}
    \label{fig:apx_by_p_comparison}
\end{figure}

Here we present results of the Homogeneous Heuristic for Parameter Setting, as well as comparisons to the transfer of parameters method outlined in Lotshaw et al. \cite{lotshaw21}, implemented using the QAOAKit software package \cite{shaydulin21_qaoakit}. Lotshaw et al. show that using one set of median (over the entire dataset at a given $n$ and $p$) parameters performs similarly to optimized parameters for each instance. Thus, we directly pull the obtained parameters from QAOAKit, which first calculates optimal parameters for all connected non-isomorphic graphs up to size $9$ at $p=1$,$2$, and $3$. For each~$p$, the median over all parameters is calculated, and these median parameters are directly applied to ten Erd\H{o}s-R\'{e}nyi graphs from $\mathcal{G}(20,1/2)$, yielding average and standard deviation of expectation values for these median parameters over the ten graphs. To compare with these transferred parameters, we display the approximation ratio achieved by parameters that are optimized with the heuristic, as described in Sec.~\ref{homogeneous_parameter_setting}, over the same ten graphs. Here the approximation ratio is defined as follows,
\begin{equation}
    \mathrm{Apx \; Ratio} = \braket{C}/c_{opt},
\end{equation}
i.e., the expected cost value returned by true QAOA, divided by the true optimal value $c_{opt}$, as determined via brute force search over all $2^n$ bitstrings. For this experiment, as well as all experiments below, the state throughout the algorithm and $\braket{C}$ are computed via full statevector simulation, even for parameters returned via the heuristic. The comparison between the  heuristic and parameter transfer is shown in Fig.~\ref{fig:apx_by_p_comparison}. 

As we can see in Fig.~\ref{fig:apx_by_p_comparison}, for low depth, the heuristic performs comparably well to parameter transfer. On an instance-by-instance basis, the approximation ratio achieved by homogeneous parameter setting minus that achieved by parameter transfer was $-.0037 \pm .0062$, $.0164 \pm .0148$, and $.0097 \pm .0183$ for $p=1$, $2$, and $3$, respectively. We do not see statistically significant differences between the two methods for any three of the depths analyzed, although the average performance of the heuristic is slightly higher in the latter two cases. This numerical evidence indicates that the method is competitive. Furthermore, the optimal parameters in the transfer case require the optimization of smaller QAOA instances, which clearly may incur some tradeoff between the size of problem one wishes to train parameters on versus the accuracy of the parameter transfer onto larger and larger instances. The parameters for comparison were pulled directly from QAOAKit data-tables, so our purpose here is not to provide a full timing comparison between the two methods. However, this demonstrates that our polynomially-scaling heuristic performs comparably with other techniques used in literature.

\subsection{Parameter Optimization at Higher Depth}\label{high_depth}

\begin{figure}[!htb]
    \centering
    \includegraphics[width=0.85\textwidth]{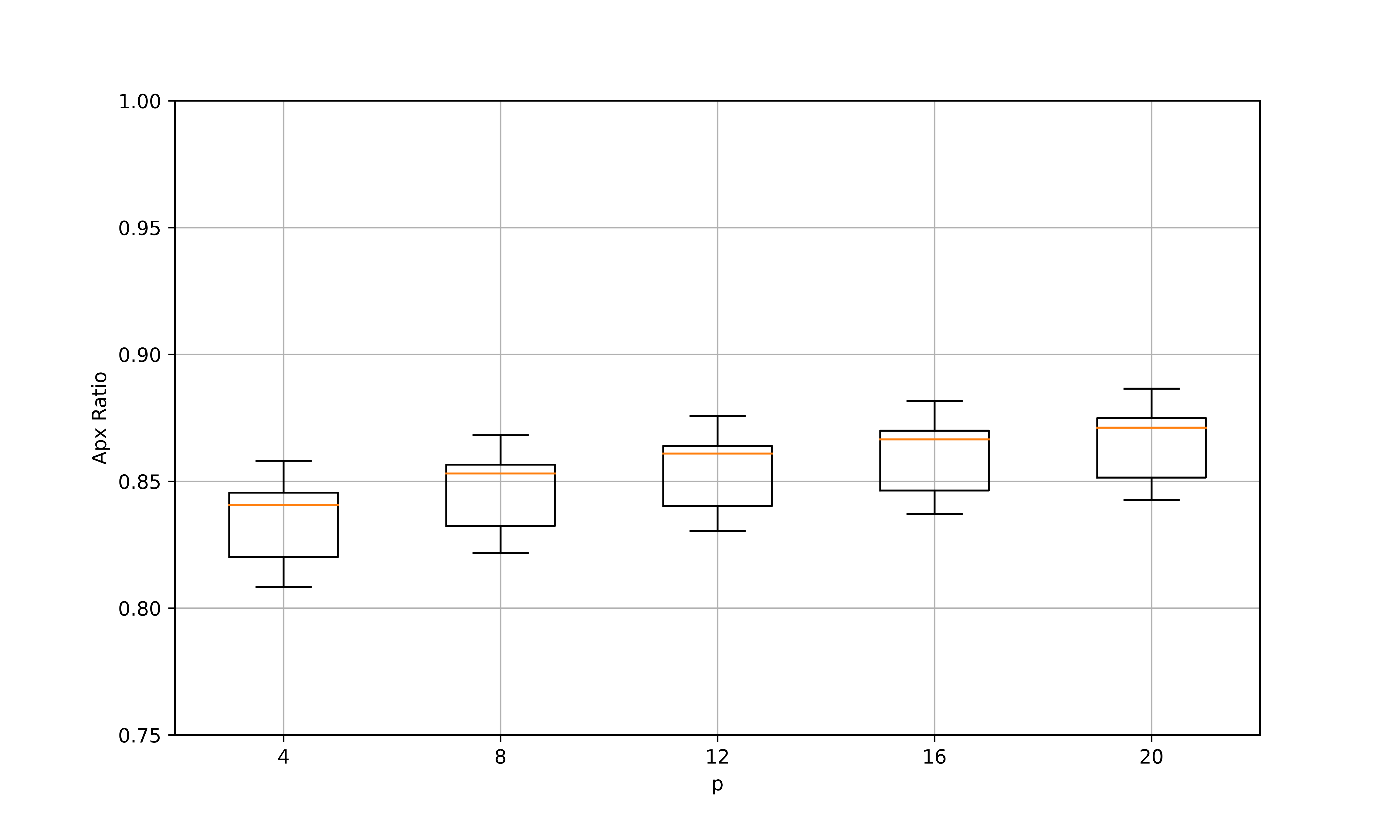}
    \caption{Box plots of approximation ratios for parameters found via homogeneous parameter setting for the same $10$ $\mathcal{G}(20,1/2)$ graphs, for $p=4$ $8$,$12$, $16$, and $20$, restricted to linear ramp schedules as described in Sec.~\ref{landscapes}.}
    \label{fig:apx_by_p_ramp_sched}
\end{figure}

To elucidate how the Homogeneous Heuristic for Parameter Setting scales with QAOA depth~$p$, we further depict box plots of the approximation ratio for a new set of ten Erd\H{o}s-R\'{e}nyi graphs $\mathcal{G}(20,.5)$ at $p$ up to $20$. For these experiments, we further restrict to linear ramp parameter schedules as described in Eq.~\eqref{eq:ramp_schedule}, to reduce the number of parameters from $2p$ to~$4$. We introduce this re-parameterization because having $2p$ free parameters, even for this relatively moderately sized $p$, results in optimization routines that do not converge in a reasonable time on the device as specified below. The results for these runs are shown in Fig.~\ref{fig:apx_by_p_ramp_sched}.

From this figure, we see that the heuristic, when implemented with linear ramp schedules, results in monotonic improvement of approximation ratios as $p$ increases. Notably, for this regime of $N=20$, $p=20$, we were not able to find previous works that efficiently returned optimized parameter schedules, even when restricted to linear ramps. Thus, these results demonstrate a regime in which the heuristic is able to return parameters that appear intractable for current devices and strategies, whether quantum, classical, or hybrid.

\textit{Numerical details}: For our simulations in this section, all calculations (excluding those pulled from the QAOAKit database) were performed using a laptop with Intel i7-10510U CPUs @ 1.80GHz and 16 GB of RAM, with no parallelization utilized. For the $20$ node graphs, all experiments clocked in under $6$ hours, where the longest times were for fully parameterized $p=3$ circuits (6 parameters). Parameters were seeded using linear ramp schedules from  \cite{zhou20} and parameter optimization was performed using the standard Broyden–Fletcher–Goldfarb–Shanno algorithm \cite{fletcher87} from the SciPy package \cite{virtanen20}.

\section{Discussion}\label{discussion}

In this work we formalized the concepts of Perfect Homogeneity and the Classical Homogeneous Proxy for QAOA. We demonstrated how to derive the necessary quantities and efficiently evaluate the proxy for combinatorial satisfaction problems with a fixed, polynomial number of randomly chosen clauses. We then provided numerical evidence to support the use of the proxy for estimating the evolution and cost expectation value of QAOA. Finally, we applied these results to construct the Homogeneous Heuristic for QAOA, and implemented this strategy for a class of MaxCut instances on graphs up to $n=20$ and $p=20$. Our results show that the heuristic on this class easily yields parameters at $p=1$, $2$, and $3$ that are comparable to those returned by parameter transfer. We further demonstrated that we are able to optimize parameters out to $p=20$ by restricting to a linear ramp schedule, obtaining the desirable property of monotonically increasing approximation ratios as the number of QAOA layers is increased. Notably, we found that the proxy seems to well-estimate both the state and cost expectation of QAOA in the particular cases when $\gamma$ remains relatively small throughout the algorithm, as well as for quantum annealing-inspired linear ramp schedules. These ramp schedules have been frequently proposed as empirically well-performing schedules \cite{zhou20,lotshaw21,wurtz21_counteradiabaticity}, which supports that the proxy may more accurately estimate QAOA expectation values for important parameter regimes and schedules of interest, even if these estimates may diverge somewhat in the 
case of arbitrarily chosen parameters. 

Several interesting research questions and future directions directly follow from our results. An immediate question is to better understand the relationship between the problem class specified, the resulting distributions $N(\cp;d,c)$ and $P(\cp)$ used for the proxy, and the effect on the parameters returned by the Homogeneous Heuristic for QAOA, especially with respect to a given problem instance to be solved. For example, a fixed instance can be drawn from a number of different possible classes, so changing the class considered can have a significant effect on the parameters returned and resulting performance. One approach to address this issue would be to extend the derivations of $N(\cp;d,c)$ and $P(\cp)$ to incorporate instance-specific information beyond just the problem class. A naive example in this vein would be to estimate the distributions via Monte Carlo sampling of bitstrings and their costs for the given instance. Furthermore, including instance-specific information appears a promising route to explicitly extending the heuristic beyond random problem classes, which can be used to study parameter schedules and performance in the worst-case setting. Finally, it is worthwhile to explore adaptations of our approach to cases where the number of unique possible costs may become large. In this case, one could imagine binning together costs close in value such that an effective cost function with much fewer possible costs is produced, and to which the proxy may be applied.

In terms of generalizing both the methods and scope of our approach, we first re-emphasize that parameter optimization for parameterized quantum circuits consists of two primary components: a parameter update scheme outer loop, and a parameter objective function evaluation subroutine. The inner subroutine is typically evaluated using the quantum computer. The key idea of our approach is to replace the inner subroutine with an efficiently computable classical strategy based on the assumption of Perfect Homogeneity. Hence a natural extension is to consider other efficiently computable proxies for the inner loop. For example, in cases where the problem instance comes with a high degrees of classical symmetries, the dimension of the effective Hilbert space can be drastically reduced, and so the evaluation and optimization of the typical parameter objective can be sped up significantly~\cite{shaydulin21_symmetries}. Similarly, different proxies may follow from related ideas and results in the literature such as the small-parameter analysis of~\cite{hadfield2021analytical}, the pseudo-Boltzmann approximation of~\cite{DiezValle22}, and classical or quantum surrogate models \cite{sung2020using,shaffer22}. We remark that a promising direction that appears relatively straightforward in light of our results is to extend the analysis of~\cite{DiezValle22} to QAOA levels beyond $p=1$. Finally, an important direction is to explicitly generalize our approach to algorithms beyond QAOA and, more generally, problems beyond combinatorial optimization, such as the parameter setting problem for Variational Quantum Eigensolvers. Generally, it is important to better understand and characterize regimes where such classical proxies are most advantages, such as when the noisy computation and measurements of real-world quantum devices is taken into account, as well as to what degree undesirable effects such as barren plateaus may apply when such proxies are utilized for parameter setting.

\section*{Acknowledgements}
We are grateful for support from NASA Ames Research Center. We acknowledge funding from the NASA ARMD Transformational Tools and Technology (TTT) Project. Part of the project was funded by the Defense Advanced Research Projects Agency (DARPA) via IAA8839 annex 114. JS, SH, TH are thankful for support from NASA Academic Mission Services, Contract No. NNA16BD14C. We also acknowledge XSEDE computational Project No. TG-MCA93S030 on  Bridges-2 at the Pittsburgh supercomputer center.

\bibliography{refs} 

\onecolumngrid

\end{document}